\documentclass[twocolumn,aps,showpacs]{revtex4}

\usepackage{color,amsmath,amsfonts,amsthm,amssymb,amscd,latexsym,graphicx, mathptmx}
\usepackage[colorlinks=true,urlcolor=webblue,linkcolor=webgreen,filecolor=webblue,citecolor=webgreen,pdfpagemode=UseOutlines,pdfstartview=FitH,pdfpagelayout=OneColumn,bookmarks=true]{hyperref}

\hypersetup{
  pdftitle=Decoherence in quantum walks on the hypercube,
  pdfauthor=Gorjan Alagic and Alexander Russell}

\def\identity{\leavevmode\hbox{\small1\kern-3.2pt\normalsize1}}

\definecolor{webgreen}{rgb}{0,.5,0}
\definecolor{webblue}{rgb}{0,0,.5}

\numberwithin{equation}{section}

\newtheorem{theorem}{Theorem}

\begin{document}

\title{Decoherence in quantum walks on the hypercube}

\author{Gorjan Alagic$^1$}
\email{alagic@math.uconn.edu}
\author{Alexander Russell$^2$}
\email{acr@cse.uconn.edu}
\affiliation{
$^1$Department of Mathematics, University of Connecticut, Storrs, CT 06269-3009\\
$^2$Department of Computer Science and Engineering, University of Connecticut, Storrs, CT 06269-2155}

\date{\today}

\begin{abstract}
  We study a natural notion of decoherence on quantum random walks
  over the hypercube.  We prove that this model possesses a
  decoherence threshold beneath which the essential properties of the
  hypercubic quantum walk, such as linear mixing times, are preserved.
  Beyond the threshold, we prove that the walks behave like their
  classical counterparts.  
\end{abstract}

\maketitle

\section{Introduction}

The notion of a \emph{quantum random walk} has emerged as an important
element in the development of efficient quantum algorithms. In
particular, it makes a dramatic appearance in the most efficient known
algorithm for element distinctness~\cite{A03}. The
technique has also provided simple separations between
quantum and classical query complexity~\cite{CCD03}, improvements
in mixing times over classical walks~\cite{NV00, MR01}, and some interesting 
search algorithms~\cite{CG04, AA05}.

The basic model has two natural variants, the \emph{continuous} model
of Childs, et al.~\cite{CFG01}, on which we will focus, and the
\emph{discrete} model introduced by Aharonov, et al.~\cite{AAKV01}. We
refer the reader to Szegedy's~\cite{S04} article for a more detailed
discussion. In the continuous model, a quantum walk on a graph $G$ is
determined by the time-evolution of the Schr\"odinger equation using
$kL$ as the Hamiltonian, where $L$ is the Laplacian of the graph and
$k$ is a positive scalar to which we refer as the ``jumping rate'' or ``energy''.
In addition to being a physically attractive model, it has been
successfully applied to some algorithmic problems as indicated above.

Such walks have been studied over a variety of graphs with special
attention given to Cayley graphs, whose algebraic structure has
provided immediate methods for determining the spectral resolution of
the linear operators that determine the system's dynamics. Once it had
been discovered that quantum random walks can offer improvement over
their classical counterparts with respect to such basic phenomena as
mixing and hitting times, it was natural to ask how robust these walks
are in the face of decoherence, as this would presumably be an issue of
primary importance for any attempt at implementation~\cite{LP03,
  SBTK03, DRKB02}.

In this article, we study the effects of a natural notion of
decoherence on the hypercubic quantum walk. Our notion of decoherence
corresponds, roughly, to independent measurement ``accidentally''
taking place in each coordinate of the walk at a certain rate $p$. We
discover that for values of $p$ beneath a threshold depending on the
energy of the system, the walk retains the basic features of the
non-decohering walk; these features disappear beyond this threshold,
where the behavior of the classical walk is recovered. 

Moore and Russell~\cite{MR01} analyzed both the discrete and the
continuous quantum walk on a hypercube. Kendon and Tregenna~\cite{KT03}
performed a numerical analysis of the effect of decoherence in the discrete case. 
In this article, we extend the continuous case with the model of decoherence 
described above. In particular, we show that up to a certain rate of
decoherence, both linear instantaneous mixing times and linear instantaneous 
hitting times still occur. Beyond the threshold, however, the walk behaves like the 
classical walk on the hypercube, exhibiting $\Theta(n \log n)$ mixing times.  As the
rate of decoherence grows, mixing is retarded by the quantum Zeno
effect.

\subsection{Results}

Consider the continuous quantum walk on the $n$-dimensional hypercube
with energy $k$ and decoherence rate $p$, starting from the initial
wave function $\Psi_0 = \vert 0 \rangle ^{\otimes n}$, corresponding 
to the corner with Hamming weight zero. We prove the following theorems 
about this walk.

\begin{theorem}
  When $p < 4k$, the walk has instantaneous mixing times at
  $$
  t_{mix} = \frac {n (2\pi c - \arccos(p^2/8k^2-1))}{\sqrt{16k^2 -  p^2}} 
  $$
  for all $c \in \mathbb{Z}$, $c > 0$. At these times, the total variation
  distance between the walk distribution and the uniform distribution is zero.
\end{theorem}

This result is an extension of the results in~\cite{MR01}, and an improvement 
over the classical random walk mixing time of $\Theta(n \log n)$. Note
that the mixing times decay with $p$ and disappear altogether
when $p \geq 4k$. Further, for large $p$, we will see that the walk is retarded 
by the quantum Zeno effect.

\begin{theorem}
When $p < 4k$, the walk has approximate instantaneous hitting times to 
the opposite corner $(1, \dots , 1)$ at times
$$
t_{hit} = \frac{2 \pi n (2c + 1)}{\sqrt{16k^2 - p^2}}
$$
for all $c \in \mathbb{Z}$, $c \geq 0$. However, the probability of
measuring an exact hit decays exponentially in $c$; the probability is
$$
P_{hit} = \left[\frac{1}{2} + \frac{1}{2}e^{-\frac{p \pi (2c + 1)}
    {\sqrt{16k^2 -  p^2}}}\right]^n\enspace.
$$
In particular, when no decoherence is present, the walk hits at 
$t_{hit} = \frac{n \pi(2c+1)}{2k},$ and it does so exactly, i.e. 
$P_{hit} = 1$. For $p \geq 4k$, no such hitting occurs.
\end{theorem}

This result is a significant improvement over the exponential hitting
times of the classical random walk, with the caveat that decoherence
has a detrimental effect on the accuracy of repeated hitting times.

Finally, we show that under high levels of decoherence,
the measurement distribution of the walk actually
converges to the uniform distribution in time $\Theta(n \log n)$,
just as in the classical case.

\begin{theorem}
For a fixed $p \geq 4k$, the walk mixes in time $\Theta(n \log n)$.
\end{theorem}

In the remainder of the introduction, we describe the continuous
quantum walk model, and recall the graph product analysis of Moore and
Russell~\cite{MR01}. In the second section, we describe our model of
decoherence, derive a superoperator that governs the behavior of the
decohering walk, and prove that it is decomposable into an $n$-fold
tensor product of a small system. We then fully analyze the small
system in the third section, and use those results to draw conclusions
about the general walk in 3 distinct regimes: $p < 4k$, $p = 4k$, and
$p > 4k$. These regimes are roughly analogous to underdamping,
critical damping, and overdamping (respectively) of a simple harmonic
oscillator with damping rate $p$ and angular frequency $2k$.

\subsection{The continuous quantum walk on the hypercube}

A continuous quantum walk on a graph $G$ begins at a distinguished
vertex $v_0$ of $G$, the initial wave function of the walk being
$\Psi_0$, where $\langle \Psi_0 \vert v \rangle = 1$ if $v = v_0$ and $0$ otherwise.
The walk then evolves according to the Schr\"odinger equation. In our
case, the graph is the $n$-dimensional hypercube. Concretely, we
identify the vertices with $n$-bit strings, with edges connecting
those pairs of vertices that differ in exactly one bit. Since the
hypercube is a regular graph, we can let the Hamiltonian $H$ be the
adjacency matrix instead of the Laplacian~\cite{GW03}; the dynamics
are then given by the unitary operator $U_t = e^{iHt}$ and the 
state of the walk at time $t$ is $\Psi_t = U_t \Psi_0$.

The following analysis makes use of the hypercube's product graph
structure; this structure will be useful again later when we consider
the effects of decoherence. The analysis below diverges from that of
Moore and Russell~\cite{MR01} only in that we allow each qubit to have
energy $k/n$ instead of $1/n$. The energy of the entire system is then
$k$.  Let
$$
\sigma_x = \left(\begin{matrix} 0 & k/n \\ k/n & 0 \end{matrix}
\right),$$
and let
$$
H = \sum_{j=1}^n \identity \otimes \cdots \otimes \sigma_x \otimes 
   \cdots \otimes \identity\enspace,
$$
where the $j$th term in the sum has $\sigma_x$ as the $j$th factor
in the tensor product. Then we have
\begin{align*}
U_t &= e^{iHt} = \prod_{j=1}^n \identity \otimes \cdots \otimes e^{it\sigma_x} \otimes \cdots
\otimes \identity = \left[e^{it\sigma_x} \right]^{\otimes n} \\
& = \left[\begin{matrix}\cos(kt/n) & i~\sin(kt/n) \\ i~\sin(kt/n) & \cos(kt/n) \end{matrix}\right]^{\otimes n}\enspace.
\end{align*}
Applying $U_t$ to the initial state $\Psi_0 = \vert 0 \rangle ^{\otimes n}$, we
have
$$
 U_t \Psi_0 = \left[ \cos\left(\frac{kt}{n}\right) \vert 0 \rangle 
  + i~\sin\left(\frac{kt}{n}\right) \vert 1 \rangle 
  \right]^{\otimes n}
  $$
  which corresponds to a uniform state exactly when $\frac{kt}{n}$
  is an odd multiple of $\frac{\pi}{4}$.

\section{A derivation of the superoperator}

We begin by recalling a model of decoherence commonly used in the
discrete model, with the intention of deriving a superoperator $U_t$,
acting on density matrices, which mimics these dynamics in our
continuous setting. The discrete model, described
in~\cite{KT03}, couples unitary evolution according to
the discrete-time quantum random walk model of Aharonov et
al.~\cite{AAKV01} with partial measurement at each step
occurring with some fixed probability $p$. Specifically, the evolution
of the density matrix can be written as
$$
\rho_{t+1} = (1-p) U \rho_t U^{\dagger} + p \sum_{i} \mathbf{P_i} 
   U \rho_t U^{\dagger} \mathbf{P_i} 
$$
where $U$ is the unitary operator of the walk, $i$ runs over the
dimensions where the decoherence occurs, and the $\mathbf{P_i}$
project in the usual ``computational'' basis~\cite{KT03}.

In the continuous setting, the unitary operator that governs the
non-decohering walk is $U_t = e^{-iHt}$, where $H$ is the normalized
adjacency matrix of the hypercube times an energy constant. To extend
the above decoherence model to this setting, recall that the
superoperator $U_t \otimes U_t^\dagger$ associated with these dynamics has the
property that
$$
\frac{d\, U_t \otimes U_t^\dagger}{dt} = i\left(e^{-iHt} \otimes e^{iHt} \right) \left[
  \identity \otimes H - H \otimes
  \identity\right]\enspace;
$$
wishing to augment these dynamics with measurement occurring at some
prescribed rate $p$, we desire a superoperator $S_t$ that satisfies
$$ 
S_{t+dt} = S_t[e^{-iHdt} \otimes e^{iHdt}][(1 - p\,dt) \identity + 
  pdt(\mathbf{P})] 
$$
where $\mathbf{P}$ is the operator associated with the
decohering measurement. Intuitively, the unitary evolution of the
system is punctuated by measurements taking place with rate $p$,
analogous to the discrete case.

Letting $e^{-iHdt} = \identity - iHdt$, 
we can expand and simplify:
\begin{align*}
S_{t+dt} & = S_t[e^{-iHdt} \otimes e^{iHdt}][(1 - pdt) \identity + pdt(\mathbf{P})]  \\
     & = S_t[(\identity - iHdt) \otimes (\identity + iHdt)][(1 - pdt) \identity + pdt(\mathbf{P})]  \\
     & = S_t[\identity \otimes \identity + idt(\identity \otimes H - H \otimes \identity) - pdt \identity \otimes \identity + pdt(\mathbf{P})]\enspace.
\end{align*}
In terms of a differential equation,
\begin{align*}
\frac{dS_t}{dt} & = \frac{S_{t+dt} - S_t}{dt}\\
    & = \frac{S_t[\identity \otimes \identity + idt(\identity \otimes H - H \otimes \identity) - pdt \left(\identity \otimes \identity + \mathbf{P}\right)] - S_t}{dt}\\
    & = S_t[i(\identity \otimes H - H \otimes \identity) - p\identity \otimes \identity + p(\mathbf{P})]\enspace.
\end{align*}
The solution is
\begin{equation}\label{superop-soln}
     S_t  = \exp\left([i(\identity \otimes H - H \otimes \identity) - p \identity \otimes \identity + p(\mathbf{P})]t\right)\enspace.
\end{equation}

We now define the decoherence operator $\mathbf P$. This operator will
correspond to choosing a coordinate uniformly at random and measuring
it by projecting to the computational basis $\{|0\rangle, |1\rangle\}.$ Let $\Pi_0$
and $\Pi_1$ be the single qubit projectors onto $\vert 0 \rangle$ and $\vert 1
\rangle$, respectively. We define
$$ 
\mathbf P = \frac{1}{n} \sum_{1 \leq i \leq n} [\Pi^i_0 \otimes \Pi^i_0 + \Pi^i_1 \otimes \Pi^i_1] 
$$
where $\Pi^i_0 = \identity \otimes \cdots \otimes \identity \otimes \Pi_0 \otimes 
\identity \otimes \cdots \otimes \identity$ with the nonidentity projector 
appearing in the $i$th place. We define $\Pi^i_1$ similarly, so that $\Pi^i_j$ 
ignores all the qubits except the $i$th one, and projects it onto $\vert j \rangle$ where 
$j \in \{0,1\}$. Note that
$$
\Pi^i_j \otimes \Pi^i_j = [\identity \otimes \identity] 
\otimes \cdots \otimes [\Pi_j \otimes \Pi_j] 
\otimes \cdots \otimes [\identity \otimes \identity]
$$
for $j \in \{0,1\}$.

\subsection{The superoperator as an $n$-fold tensor product}

The pure continuous quantum walk on the $n$-dimensional hypercube is
easy to analyze, in part, because it is equivalent to a system of $n$
non-interacting qubits. We now show that, with the model of
decoherence described above, each dimension still behaves
independently. In particular, the superoperator that dictates the
behavior of the walk is decomposable into an $n$-fold tensor product.

Recall the product formulation of the non-decohering Hamiltonian
$$
H = \sum_{j=1}^n \identity \otimes \cdots \otimes \sigma_x 
   \otimes \cdots \otimes \identity 
$$
where
$$
\sigma_x = \left(\begin{matrix} 0 & k/n \\ k/n & 0 \end{matrix} \right)
$$
with $\sigma_x$ appearing in the $j$th place in the tensor
product. We have given each single qubit energy $k/n$, resulting
in a system with energy $k$. This choice will allow us to 
precisely describe the behavior of the walk in terms of the 
relationship between the energy of the system and the rate of
decoherence.

We can write each of the terms in the exponent of the
superoperator from (\ref{superop-soln}) as follows:
\begin{align*}
\identity \otimes H &= \sum_{j=1}^n [\identity \otimes \identity] \otimes \cdots \otimes [\identity \otimes \sigma_x] \otimes \cdots \otimes [\identity \otimes \identity]\enspace, \\
H \otimes \identity &= \sum_{j=1}^n [\identity \otimes \identity] \otimes \cdots \otimes [\sigma_x \otimes \identity ] \otimes \cdots \otimes [\identity \otimes \identity]\enspace.
\end{align*}
Our decoherence operator can also be written in this form:
\begin{eqnarray*}
 {\mathbf P} &=& \frac{1}{n} \sum_{j=1}^n [\Pi^i_0 \otimes \Pi^i_0 + \Pi^i_1 \otimes \Pi^i_1] \\
             &=& \frac{1}{n} \sum_{j=1}^n ([{\identity} \otimes {\identity}] \otimes \cdots \otimes [\Pi_0 \otimes \Pi_0] \otimes \cdots \otimes [{\identity} \otimes {\identity}]\\
                                      &&+ [{\identity} \otimes {\identity}] \otimes \cdots \otimes [\Pi_1 \otimes \Pi_1] \otimes \cdots \otimes [{\identity} \otimes {\identity}])\enspace.
\end{eqnarray*}
The identity operator has a consistent decomposition: $\identity \otimes
\identity = \frac{1}{n}\sum_{j=1}^n [{\identity} \otimes {\identity}] \otimes \cdots \otimes
[{\identity} \otimes {\identity}].$ We can now put these pieces together to
form the superoperator:
\begin{align*}
S_t  
& = \exp\left(it(\identity \otimes H) - it(H \otimes \identity) - pt \identity \otimes \identity + pt\mathbf{P}\right) \\
& = \exp\left(\sum_{j=1}^n [{\identity} \otimes {\identity}] \otimes \cdots \otimes \mathbf{A} \otimes \cdots [{\identity} \otimes {\identity}]\right) \\
& = \prod_{j=1}^n [{\identity} \otimes {\identity}] \otimes \cdots \otimes e^\mathbf{A} \otimes \cdots [{\identity} \otimes {\identity}]\\
& = \left[e^\mathbf{A}\right]^{\otimes n}
\end{align*}
where
\begin{eqnarray*}
 \mathbf{A} 
&=& \frac{t}{n}[(\identity \otimes in\sigma_x) - (in\sigma_x \otimes \identity) - p({\identity} \otimes {\identity})\\
   &&+ p(\Pi_1 \otimes \Pi_1) + p(\Pi_0 \otimes \Pi_0)] \\
&=& \frac{t}{n}\left( \begin{matrix} 0   &  ik & -ik &  0 \\
                        ik  & -p  &  0  & -ik \\
                        -ik &  0  & -p  &  ik \\
                        0   & -ik &  ik &  0 \\
    \end{matrix} \right).
\end{eqnarray*}
Notice that for $p = 0$, $\left[e^{\mathbf{A}}\right]^{\otimes n}
= \left[e^{-it\sigma_x} \otimes e^{it\sigma_x}\right]^{\otimes n}$, 
which is exactly the superoperator formulation of the dynamics of 
the non-decohering walk. 

\section{Small-system behavior and analysis of the walk}

So far we have shown that the walk with decoherence is still
equivalent to $n$ non-interacting single-qubit systems. We now analyze
the behavior of a single-qubit system under the superoperator
$e^\mathbf{A}$. The structure of this single particle walk will allow us to then
immediately draw conclusions about the entire system. 

The eigenvalues of $\mathbf{A}$ are $0$, $- \frac{pt}{n}$, $\frac{-p t
  - \alpha t}{2n}$ and $\frac{-p t + \alpha t}{2n}$.  Here $\alpha =
\sqrt{p^2-16k^2}$ is a complex constant that will later turn out to be
important in determining the behavior of the system as a function of
the rate of decoherence $p$ and the energy $k$. The matrix exponential
of $\mathbf{A}$ in this spectral basis can then be computed by
inspection. To see how our superoperator acts on a density matrix
$\rho_0$, we may change $\rho_0$ to the spectral basis, apply the diagonal
superoperator to yield $\rho_t$, and finally change
$\rho_t$ back to the computational basis.  At that point we can apply the
usual projectors $\Pi_0$ and $\Pi_1$ to determine the probabilities of
measuring $0$ or $1$ in terms of time.

\begin{figure}[h]
\centering
\includegraphics[width=3in]{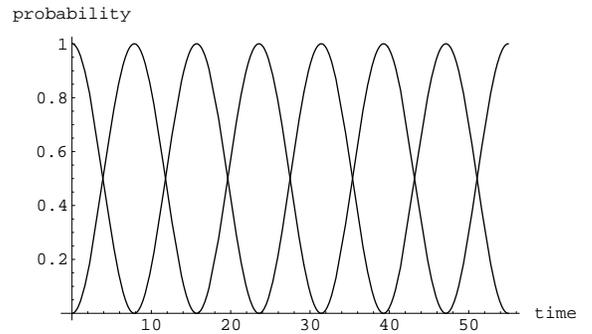}
\caption{The $p=0$ case - no decoherence: a plot of $P[0]$ and $P[1]$ versus time, for $k =1$, $n=5$, $p = 0$}
\label{fig1}
\end{figure}

Let $\Psi_0 = \vert 0 \rangle$ and $\rho_0 = \vert \Psi_0 \rangle 
\langle \Psi_0 \vert$. In the diagonal basis,
$$ 
\rho_0 = \left[\begin{matrix} 1/2 \\ 0 \\ \frac{1}{4}(-1 + \frac{p}{\alpha}) \\ \frac{1}{4}(-1 - \frac{p}{\alpha}) \end{matrix} \right] 
$$
and thus at time $t$ we have
$$
\rho_t = e^{\mathbf{A}}\rho_0 = \left[\begin{matrix} 1/2 \\ 0 \\
            \frac{1}{4}e^{\frac{-p t - \alpha t}{2n}}(-1 + \frac{p}{\alpha}) 
            \\ \frac{1}{4}e^{\frac{-p t + \alpha t}{2n}}(-1 - \frac{p}{\alpha}) \end{matrix} \right].
$$
If we then change back to the computational basis and project by 
$\Pi_0$ and $\Pi_1$, we may compute the probabilities of measuring 
$0$ and $1$ at a particular time $t$:
$$ 
P[0] = \frac{1}{4}\left[2 + e^{\frac{-p t - \alpha t}{2n}}(1 - p/ \alpha) 
        + e^{\frac{-p t + \alpha t}{2n}}(1 + p/ \alpha)\right] 
$$
$$ 
P[1] = \frac{1}{4}\left[2 - e^{\frac{-p t - \alpha t}{2n}}(1 - p/ \alpha) 
                             - e^{\frac{-p t + \alpha t}{2n}}(1 + p/ \alpha)\right] 
$$
which can be simplified somewhat to
$$
P[0] = \frac{1}{2} + \frac{1}{2} e^{-\frac{pt}{2n}} \left[\cos\left(\frac{\beta t}{2n}\right) + 
                         \frac{p}{\beta}~\sin\left(\frac{\beta t}{2n}\right) \right] 
$$
$$
P[1] = \frac{1}{2} - \frac{1}{2} e^{-\frac{pt}{2n}} \left[\cos\left(\frac{\beta t}{2n}\right) + 
                         \frac{p}{\beta}~\sin\left(\frac{\beta t}{2n}\right) \right].
$$
Here we have let $\beta = -i \alpha = \sqrt{16k^2-p^2}$ for simplicity. A quick
check shows that when $p = 0,$ $P[0] =
\cos^2\left(\frac{kt}{n}\right)$ and $P[1] =
\sin^2\left(\frac{kt}{n}\right)$, which are exactly the dynamics of
the non-decohering walk. The probabilities for this non-decohering
case are shown in Figure~\ref{fig1}.

The three regimes mentioned before are immediately apparent. 
For $p < 4k$, $\beta$ is real. When $p = 4k$, we have $\beta = 0$, 
which appears to be a serious problem at first glance. Finally, 
for $p > 4k$, $\beta$ is imaginary. We now address each of these
three situations in detail.

\subsection{The case $p < 4k$ : linear mixing and hitting times}

\begin{figure}[b]
\centering
\includegraphics[width=3in]{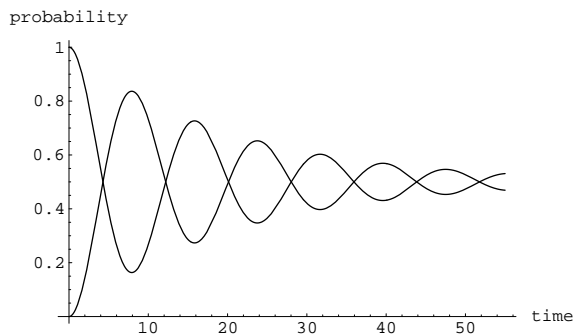}
\caption{The $p<4k$ case: a plot of $P[0]$ and $P[1]$ versus time, for $k =1$, $n=5$, $p = 0.5$}
\label{fig2}
\end{figure} 

When $p < 4k$, we recover the perhaps most interesting feature of the
non-decohering walk: the instantaneous mixing time is linear in $n$.
To exactly determine the mixing times for our decohering walk, we
solve $P[0] = P[1] = \frac{1}{2}$; this amounts to determining when 
$$
\gamma
= \frac{1}{2} e^{-\frac{pt}{2n}} \left[\cos\left(\frac{\beta t}{2n}\right)
  + \frac{p}{\beta}~\sin\left(\frac{\beta t}{2n}\right) \right]
$$
equals zero. Clearly the exponential decay term results in mixing as $t \to \infty$; our
principle concern, however, is with the periodic mixing times
analogous to those of the original walk. We thus ignore the
exponential term when solving the equality $\gamma = 0$, which yields
$$ 
\frac{p^2}{\beta^2} = \frac{1 + \cos(\beta t/n)}{1 - \cos(\beta t/n)}.
$$
This equation actually has more solutions than the one we
started with, because of the use of half-angle
formulas for simplification. The solutions that we want are
$$ 
t_{mix} = \frac {n}{\beta} \left[2\pi c - \arccos\left(\frac{p^2}{8k^2}-1\right) \right] 
$$
where $c$ ranges over the positive integers. Evidently, the
mixing times still occur in linear time; an example is shown in Figure~\ref{fig2}. 
Note also that if we let $p = 0$, we have $t_{mix} = n\pi(2c - 1)/(4k)$, which are exactly the nice periodic 
mixing times of the non-decohering walk. In the decohering case,
however, these mixing times drift towards infinity, and cease 
to exist altogether beyond the threshold of 
$p = 4k$. This proves Theorem 1.

We now wish to determine when our small system is as close as possible
to $\vert 1 \rangle$. Since our large-system walk begins at $\vert 0 \rangle^{\otimes
  n}$, this will correspond to approximate hitting times to the
opposite corner $\vert 1 \rangle^{\otimes n}$. These times correspond to local
maxima of $P[1]$; the solutions are
$$
t_{hit} = 2n \pi \left( \frac{2c + 1}{\beta} \right)
$$
where $c$ ranges over the non-negative integers. At these 
points in time, the value of $P[1]$ is
$$ 
\frac{1}{2} + \frac{1}{2}e^{-(2c+1)\frac{p \pi}{\beta}} 
$$
which immediately yields Theorem 2.

\subsection{The breakpoint case $p = 4k$}

We first observe that $t_{mix} \to \infty$ as $p \to 4k$. Hence,
we do not expect to see any mixing in this case. To analyze 
the probabilities exactly, we take the limit of $\gamma$ as
$p \to 4k$. The solution is
\begin{equation}\label{pequals4k}
\lim_{p \to 4k} \gamma = \frac{1}{2}e^{-\frac{2kt}{n}}\left[1 + \frac{2kt}{n} \right]\enspace.
\end{equation}
Indeed, since $k$, $t$ and $n$ are all positive, $\gamma$ is 
zero only in the limit as $t \to \infty$. The linear 
mixing and hitting behavior from the previous section has 
entirely disappeared. As in the critical damping of simple 
harmonic motion, a small decrease in the rate $p$ can result 
in drastically different behavior, in this case a return to 
linear mixing and hitting. We leave the limiting mixing 
analysis of this case for the next section, where we develop 
some relevant tools.

\subsection{The case $p > 4k$ and the limit to the classical walk}

\begin{figure}[h]
\centering
\includegraphics[width=3in]{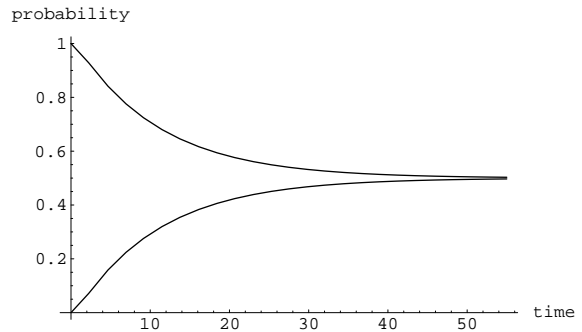}
\caption{The $p>4k$ case: a plot of $P[0]$ and $P[1]$ versus time, for $k=1$, $n=5$, $p = 9$}
\label{fig3}
\end{figure}

The goal of this section is to show two interesting consequences of
the presence of substantial decoherence in the quantum walk on the
hypercube.  First, we will show that for a fixed $p \geq 4k$, the walk
behaves much like the classical walk on the hypercube, mixing in time
$\Theta(n \log n)$. Second, we show that as $p \to \infty$, the walk suffers from
the quantum Zeno effect. Informally stated, the rate of decoherence is
so large that the walk is continuously being reset to the
initial wave function $|0\rangle^{\otimes n}$ by measurement.

\subsubsection{Recovering classical behavior}

Consider a single qubit. Let $P$ be the distribution obtained by full
measurement at time $t$, and $U$ the uniform distribution:
$$ 
P(0) = \frac{1}{2}+ \gamma, \qquad P(1) = \frac{1}{2}- \gamma, \qquad U(0) = U(1) = \frac{1}{2},
$$
where
$$ 
\gamma = \frac{1}{4}\left[e^{\frac{(-p-\alpha)t}{2n}}(1-p/\alpha) + e^{\frac{(\alpha-p)t}{2n}}(1+p/\alpha)\right].
$$
For $x = (x_1, \dots, x_n) \in \mathbb{Z}_2^n$,  
$$
P^n(x) = \prod_{i=1}^n P(x_i) \qquad\text{and}\qquad U^n(x) = 2^{-n}
$$
are the analogous product distributions in the $n$-dimensional case.
To analyze the limiting mixing behavior of the walk, we will 
consider the total variation distance $\|P^n - U^n\| = 
\sum_x |P^n(x) - U^n(x)|$ between these distributions. In 
order to give bounds for total variation, we will use 
\emph{Hellinger distance}~\cite{ASYMP}, defined as
follows:
$$
H(A, B)^2 = \sum_{x} \left(\sqrt{A(x)} - \sqrt{B(x)}\right)^2 = 1 - \sum_{x}\sqrt{A(x)B(x)}.
$$
We will make use of the following two properties of Hellinger
distance:
$$
1 - H(A^n, B^n)^2 = (1-H(A,B)^2)^n\enspace,
$$
and
\begin{equation}
  \|A - B\| \leq 2H(A, B) \leq 2\|A - B\|^{1/2}.
  \label{helltv}
\end{equation}
The first property makes it easy to work with product
distributions. The second gives a nice relationship between
Hellinger distance and total variation distance. In our case, 
\begin{align*}
  H(P^n,U^n)^2 
& = 1 - (1-H(P,U)^2)^n \\
& = 1 - \left(\frac{1}{2}\sqrt{1+2\gamma} + \frac{1}{2}\sqrt{1-2\gamma}\right)^n\\
& = 1 - \left(1 - \frac{\gamma^2}{2} + O(\gamma^3)\right)^n.
\end{align*}
And hence, by (\ref{helltv}),
$$
\|P_n - U_n\|^2 \leq 4 - 4\left(1 - \frac{\gamma^2}{2} + O(\gamma^3)\right)^n.
$$
Consider the walk with decoherence rate $p > 4k$. We have
$\alpha = \sqrt {p^2 - 16k^2} < p$, where $\alpha$ and $p$ are
positive and real. It follows that for a fixed
$p > 4k$, $\gamma \to 0$ and $\|P^n - U^n\| 
\to 0$ as $t \to \infty$. Hence the walk does indeed mix eventually, and the measurement 
distribution in fact converges to the uniform distribution. Let $t = d \cdot n\log n$ where $d > 0$
is a constant, and rewrite $\gamma$ as follows:
$$ 
\gamma = \frac{1}{4}e^{-(p-\alpha)\frac{d \log n}{2}}\left[(1-p/\alpha) + e^{\frac{-\alpha d \log n}{2}}(1+p/\alpha)\right]. 
$$
Suppose we choose $d$ such that $d > (p - \alpha)^{-1}$.
Then $\gamma = o(n^{-1/2})$, which implies that $\|P^n - U^n\| = o(1)$. On the 
other hand, if $d < (p - \alpha)^{-1}$, then $\gamma = \omega(n^{-1/2})$ and 
there exists a constant $\epsilon$ such that $\|P^n - U^n\| \geq \epsilon > 0$. 
This shows that the walk mixes in time $\Theta(n \log n)$ when $p > 4k$. Notice that when 
$p = 4k$, $(p - \alpha)^{-1} = (4k)^{-1}$, so that the same 
technique easily extends to that case via equation 
(\ref{pequals4k}). This completes the proof of Theorem 3.

\subsubsection{Quantum Zeno effect for large $p$}

Recall from the previous section that the time required to mix when
$p > 4k$ is
$$
t \geq \frac{n~ \log n}{p - \alpha}
$$
which clearly increases with $p$. Further, for large $p$, $p/\alpha$
tends to $1$, and hence $\gamma$ tends to $1/2.$ Notice that $\gamma = 1/2$
corresponds to remaining at the initial state forever. We conclude that the 
mixing of the walk is retarded by the quantum Zeno effect, where measurement occurs 
so often that the system tends to remain in the initial state.

\begin{acknowledgments}
Alexander Russell gratefully acknowledges the support of the National Science Foundation, 
under the grants CAREER CCR-0093065, CCR-0220264, EIA-0218443, and ARO grant W911NF-04-R-0009.
The authors are grateful to Viv Kendon for helpful suggestions.
\end{acknowledgments}

\end{document}